\documentclass[final,5p,times]{elsarticle}

\usepackage{graphicx}

\usepackage{amssymb}
\usepackage{amsmath}

\usepackage[latin1]{inputenc}
\usepackage{hyperref}
\usepackage{color}
\usepackage[all]{hypcap}

 \def\be{\begin{equation}}
 \def\ee{\end{equation}}
 \def\bea{\begin{eqnarray}}
 \def\eea{\end{eqnarray}}
 \def\bean{\begin{eqnarray*}}
 \def\eean{\end{eqnarray*}}
 \def\gsim{\mathrel{\rlap{\lower0.2em\hbox{$\sim$}}\raise0.2em\hbox{$>$}}}
 \def\ksim{\mathrel{\rlap{\lower0.2em\hbox{$\sim$}}\raise0.2em\hbox{$<$}}}

\begin{document}

\begin{frontmatter}


\title{Collisional energy loss of heavy quarks}
\author[fr]{Alex Meistrenko}
\ead{meistren@uni-frankfurt.de}
\author[fr]{Jan Uphoff}
\author[fr]{Carsten Greiner}
\author[ct]{Andr\'e Peshier}
\address[fr]{Institut f\"ur Theoretische Physik, Johann Wolfgang Goethe-Universit\"at Frankfurt,
Max-von-Laue-Str.~1, D-60438 Frankfurt am Main, Germany}
\address[ct]{University of Cape Town, Physics Department, Rondebosch 7700, South Africa}


\begin{abstract}
We develop a transport approach for heavy quarks in a quark-gluon plasma, which is based on improved binary collision rates taking into account quantum statistics, the running of the QCD coupling and an effective screening mass adjusted to hard-thermal loop calculations.
We quantify the effects of in-medium collisions by calculating the heavy flavor nuclear modification factor and the elliptic flow for RHIC energies, which are comparable to radiative effects.
We also derive an analytic formula for the mean collisional energy loss of an energetic heavy quark in a streaming quark gluon plasma.
\end{abstract}


\end{frontmatter}


\section{Introduction}
\label{chap:100}
Jet quenching, as observed in heavy ion collisions at RHIC \cite{Adams:2005dq,Adcox:2004mh,Arsene:2004fa,Back:2004je}, has become an important probe of a deconfined state of matter with quasi-free quarks and gluons, the so-called quark gluon plasma (QGP). Indeed, this observable allows inferences on the properties of the expanding medium. Especially jet quenching of heavy flavors charm $(c)$ and bottom $(b)$ could be an ideal probe of the QGP. Due to the large mass\footnote{$M_c\sim 1.3\,$GeV, $M_b\sim 4.6\,$GeV} of these quarks, $M \gg T$, they are produced in a very early stage of the collision and can later be considered as test particles in the (equilibrated) heat bath.

The quenching of light quarks is well understood as an effect of radiative energy loss when propagating through the me\-dium \cite{Baier:1996sk,Gyulassy:2000fs}. However, in the case of heavy quarks pure radiative energy loss seems not sufficient to explain experimental data (mainly from electron yields) on heavy flavor quenching \cite{Abelev:2006db,Adare:2006nq,Bielcik:2005wu,Wicks:2005gt}, suggesting
surprisingly strong attenuation. There is an ongoing debate on the underlying reasons,
which could be possibly clarified by analyzing spectra from heavy flavor decays. An additional contribution of suppression, potentially important for the range of smaller (to medium) momenta, is the collisional energy loss by elastic processes as discussed in \cite{Wicks:2005gt,Mustafa:2003vh,Moore:2004tg} and with further modifications in \cite{Peshier:2008bg, Gossiaux:2008jv, Gossiaux:2009mk, vanHees:2005wb, vanHees:2007me, Uphoff:2011ad}. It will turn out that in our approach a fairly large collisional energy loss is predicted, although some comparisons between collisional and radiative contributions \cite{Qin:2007rn,Zakharov:2007pj} indicate another picture where the collisional loss should be relatively small ($\sim 20\%$) compared to the radiative one. This conclusion was, however, based on a fixed coupling approach, and using a naive approximation for the required infrared regulator in the QCD cross section. In the following work both aspects will be considered in a more precise way, aiming at providing a computational framework to describe binary parton collision effects more reliably, which seems necessary as a baseline to understand quenching in heavy-ion experiments. Noted in this context is that certain observables might be more sensitive to the inclusion of binary collisions than the averaged energy loss, due to different probability distributions of radiative and collisional processes \cite{Peshier:2008bg,Qin:2007rn}.

\section{Mean energy loss with QGP flow}
\label{chap:200}
We start with a short overview of the collisional energy loss of a heavy quark in a static QGP with temperature $T$. The average loss per unit length $dE/dx$ has recently been recalculated \cite{Peigne:2007sd,Peigne:2008nd}, revising a previous result by Braaten and Thoma \cite{Braaten:1991jj,Braaten:1991we} in the large-energy limit. In general, the mean energy loss of a heavy quark, with velocity $v=p/E$, due to binary interactions with light/strange quarks and gluons ($i\in\left\{u \mbox{ or } d,s,g\right\}$) from the plasma,
\be
\frac{dE_i}{dx}=\frac{1}{v}\int d\omega\,\frac{\partial\Gamma_i}{\partial\omega}\omega\,,
\label{eqn:201}
\ee
is the first moment of the differential transition rate $\partial\Gamma_i/\partial\omega$ with respect to the energy transfer $\omega=E-E'$ in the interaction. To calculate $dE/dx$ beyond leading-log accuracy (see Eq.~\eqref{eqn:204} below), it is convenient to decompose the phase space of the invariant momentum transfer $t$ into a soft and a hard part with respect to a separation scale $|t^*|$.
In the weak coupling limit, $\alpha \ll 1$,  $|t^*|$ can be chosen arbitrarily in the range $\alpha T^2 \ll |t^*|\ll T^2$. A sufficient separation of the `soft' screening scale, which is of the order of the Debye mass $m_D \sim \sqrt\alpha T$, and the `hard' thermal scale, $\sim T$, seems to persist (at least for the observables under consideration) even for moderately large coupling, as will be discussed below.

The soft contribution to $dE/dx$ can be calculated from the imaginary part of the hard thermal loop (HTL) dressed heavy quark self-energy \cite{Peigne:2007sd,Peigne:2008nd}. The hard contribution can be obtained from the interaction rate evaluated with tree-level matrix elements $\mathcal M_i$, with a constrained momentum transfer $|t| \ge |t^*|$,
\be
\begin{aligned}
\frac{dE_i^{(h)}}{dx}=&\frac{1}{2Ev}\int_k\frac{n_i(k)}{2k}\int_{k'}\frac{\overline{n}_i\left(k'\right)}{2k'}\int_{p'}\frac{1}{2E'} \, \Theta(|t|-|t^\star|)\, \times\\
&\times\left(2\pi\right)^4\delta^{(4)}\left(P+K-P'-K'\right)\frac{1}{d}\sum_{spin, color}\left|\mathcal M_i\right|^2\omega\,,
\end{aligned}
\label{eqn:202}
\ee
where $n_i(k) = (\exp(k/T) \mp 1)^{-1}$ denotes the Bose or Fermi distribution\footnote{We note that we may approximate the distribution functions by the massless ideal gas expressions even for moderately strong coupling, since the particle densities are closely related to the entropy density which is known, from lattice QCD calculations, to deviate from the free limit only by ${\cal O}(10\%)$ even fairly close to the transition/crossover temperature.}
of the thermal target particle $i$, similarly $\overline{n}_i(k') = 1 \pm n_i(k')$ for the final states, and we use the short-hand notation $\int_k:=\int d^3k/(2\pi)^3$. $P$ and $K$ ($P'$and $K'$) denote the initial (final) 4-momenta of the collisional partners; $P = (E,\vec p)$ and $P'$ are the heavy quark momenta. The sum of the matrix elements $|\mathcal M_i|^2$ runs over all initial and final states and is divided by the degeneracy factor $d=6$ of the heavy quark.

In the large-energy limit, for $E\gsim M^2/T$, and to the level of accuracy we are interested in, Eq.~\eqref{eqn:202} can be simplified \cite{Peigne:2007sd,Peigne:2008nd},
\be
\frac{dE_i^{(h)}}{dx}=d_i\int_k\frac{n_i(k)}{2k}\int_{t_{\rm min}}^{t^*}dt\, (-t)\frac{d\sigma_i}{dt}\, .
\label{eqn:203}
\ee
Here $d\sigma_i/dt$ are the heavy quark Born cross sections \cite{Combridge:1978kx} for the interaction with medium particles of type $i$ and degeneracy factor $d_i$, and $t_{\rm min} = -(s-M^2)^2/s$ in terms of the invariant energy $s = (K+P)^2$. Keeping sub-leading terms in the calculation of the soft and hard phase space contributions leads to the NLL (next-to-the leading energy logarithm) formula \cite{Peigne:2007sd,Peigne:2008nd}
\be
\frac{dE}{dx}=\frac{4\pi\alpha_s^2T^2}{3}\left[\left(1+\frac{n_f}{6}\right)\ln\left(\frac{ET}{m_D^2}\right)+\frac{2}{9}\ln\left(\frac{ET}{M^2}\right)+c(n_f)+\ldots\right]\,.
\label{eqn:204}
\ee
The constant $c(n_f)$ depends on the number of active flavors, in the present context $n_f=3$ ($u,d,s$),
\be
\begin{aligned}
c(n_f)&=a\cdot n_f+b\simeq 0.146\cdot n_f+0.050 \, ,\\
a&=\frac{2}{3}\ln\left(2\right)-\frac{1}{8}+\frac{1}{6}c \, ,\\
b&=\frac{31}{9}\ln\left(2\right)-\frac{101}{108}+\frac{11}{9}c \, ,\\
c&=\frac{\zeta'\left(2\right)}{\zeta\left(2\right)}-\gamma\simeq -1.147 \, ,\\
\gamma&\simeq 0.577\,\mbox{(Euler's constant).}
\end{aligned}
\label{eqn:205}
\ee
The first (Coulomb) logarithm in Eq.~\eqref{eqn:204} arises from $t$-channel interactions and was first obtained by Bjorken \cite{Bjorken:1982tu}. The second logarithm, due to QCD Compton scattering where a potential collinear divergence is screened by the heavy quark mass, was missed for some time in the literature.

We point out that formula \eqref{eqn:204} assumes a {\em fixed} coupling $\alpha_s$. Taking into account the momentum dependence of the QCD coupling amounts to evaluating prefactors of $\alpha_s$ in \eqref{eqn:204} at scales specific for each contribution: for the Coulomb-log term $\alpha^2_s\rightarrow \alpha(ET)\alpha(m_D^2)$ \cite{Peshier:2006hi}, and $\alpha^2_s\rightarrow \alpha(ET)\alpha(M^2)$ in the Compton logarithm term \cite{Peigne:2008nd}.

\begin{figure}
 \includegraphics[width=8.5cm,height=6.1cm]{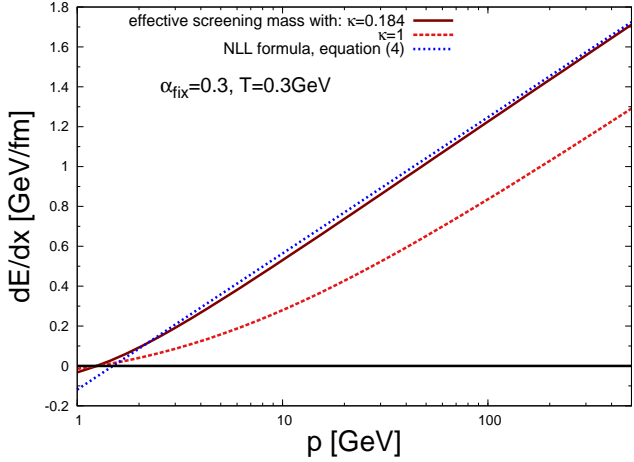}
 \centering
 \caption{Mean energy loss of a heavy quark, in the fixed-coupling approximation, as a function of its momentum for two different values of $\kappa$ in comparison to the analytic NLL-formula \eqref{eqn:204}; $\kappa = 0.184$ is the adjusted parameter \eqref{eqn:207}, $\kappa = 1$ corresponds to screening with the Debye mass.}
 \label{fig:201}
\end{figure}

In the numerical simulation of the next section we will use differential transition rates calculated from the QCD Born cross sections integrated over the whole available phase space, however with the singular $t$-channel contributions screened by a mass term, i.\,e.\ we replace $d\sigma^{[t]}/dt \sim t^{-2} \;\to\; (t-\mu^2)^{-2}$. The screening mass $\mu$ has often been chosen {\em ad hoc} to be equal to the Debye mass -- although they have only the same {\em parametric} dependence on the coupling, $\mu^2 \sim \alpha T^2 \sim m_D^2$. We illustrate this point by showing in Fig.~\ref{fig:201} that the resulting energy loss would not converge to the asymptotic formula \eqref{eqn:204}.
In order to match this analytic result one actually has to introduce an effective screening mass of the form \cite{Peshier:2008bg}
\be
\mu^2(t)=\kappa\cdot 4\pi\left(1+\frac{n_f}{6}\right)\alpha\left(t\right)T^2\,,
\label{eqn:206}
\ee
which $i)$ depends on the momentum exchange via the running coupling, and $ii)$ also differs by a prefactor from the Debye mass
\be
	m_D^2(T)
	=
	4\pi\left(1+\frac{n_f}{6}\right)\alpha(m_D^2)\,T^2 \, .
	\label{eqn:mD2}
\ee
This prefactor $\kappa$ has been determined \cite{Peshier:2008bg,Gossiaux:2008jv},
\be
\kappa=\frac{1}{2e}\simeq 0.184\,.
\label{eqn:207}
\ee
Let us comment at this point on the separation of `soft' and `hard' scales, $\sim \alpha T^2$ and $\sim T^2$ respectively (which underlies the derivation of equation \eqref{eqn:204}), for larger coupling. Then prefactors become important; for the typical hard scale the thermal average reads $\langle k^2 \rangle \approx 10T^2$. This value happens to coincide approximately with an upper bound, seen in non-perturbative lattice QCD calculations,\footnote{We underline that the implicit relation \eqref{eqn:mD2}, where $m_D$ sets the scale of the running coupling \cite{Peshier:2006ah}, is in quantitative agreement with lattice QCD calculations even fairly close to the transition temperature.} for the squared Debye mass, $m_D^2 \ksim 10T^2$.
Therefore the facts that $\kappa \ll 1$ in the cut-off \eqref{eqn:206} and the running coupling in \eqref{eqn:206} being smaller than $\alpha(m_D^2)$ in the Debye mass \eqref{eqn:mD2} for harder momentum exchanges, suggest that a certain separation of scales persists even for moderately large coupling, and that our approach may provide reasonable numerical estimates.

\medskip
We now generalize the mean energy loss formula \eqref{eqn:204} to the case of a QGP heat bath with a collective flow. In this case the thermal distribution functions $n_i(k)$ have to be replaced by the J\"uttner functions for bosons and fermions, which depend on the velocity $\vec\beta$ of the medium,
\be
	n_J\left( \vec k \right)
	=
	\frac1{e^{\gamma\left(k-\vec k \cdot\vec\beta\right)/T} \mp 1}\, ,
	\label{eqn:208}
\ee
where $\gamma = (1-\vec\beta\,^2)^{-1/2}$.
Noting that the RHS of \eqref{eqn:203}, with J\"uttner distributions in place, is a manifestly covariant expression, we expect that the energy loss in a streaming QGP can be obtained from \eqref{eqn:204} by simply replacing $E\rightarrow p_\mu\beta^\mu=\gamma\left(E-\vec p\cdot\vec\beta\right)$ in the two logarithms, i.\,e.\footnote{To take into account the running of the QCD coupling, the prefactor $\alpha_s^2$ in \eqref{eqn:216} is to be replaced as discussed below Eq.~\eqref{eqn:204}, however now with $\alpha_s(p_\mu \beta^\mu T)$ instead of $\alpha_s(ET)$.\label{footnote: run coupling}}
\be
\frac{dE}{dx}=\frac{4\pi\alpha_s^2T^2}{3}\Bigg[\left(1+\frac{n_f}{6}\right)\ln\frac{p_{\mu}\beta^{\mu}T}{m_D^2}
+\frac{2}{9}\ln\frac{p_{\mu}\beta^{\mu}T}{M^2} + c(n_f) + \ldots \Bigg] \, ,
\label{eqn:216}
\ee
which we will prove in appendix \ref{appendix:100} for the Coulomb term.
By covariance we also expect that the constant $c(n_f)$ is the same as in the static case \eqref{eqn:205} (where it is $E$-independent) -- which we will justify numerically in Section 3 (a rigorous proof would require a more cumbersome evaluation of thermal J\"uttner integrals both for the hard and the soft contributions, which is not investigated here).
\begin{figure}
 \includegraphics[width=8.5cm,height=6.1cm]{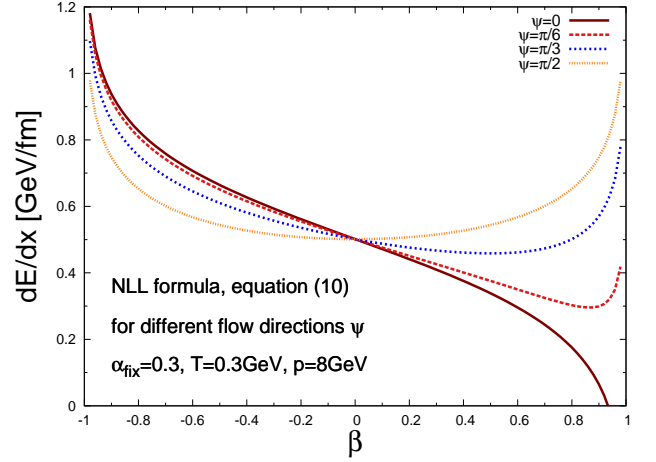}
 \centering
 \caption{Mean energy loss \eqref{eqn:216} of a $p = 8\,$GeV charm quark as a function of the medium velocity $\beta$ for different flow directions of the QGP with respect to the charm momentum vector, in fixed-coupling approximation.}
 \label{fig:202}
\end{figure}
We illustrate in Fig.~\ref{fig:202} that flow of the QGP, as expected in heavy-ion experiments, can have sizable effects on the mean collisional energy loss of heavy quarks.

\section{Monte Carlo simulation with quantum transition rates}
\label{chap:300}
The following simulations are based on the heavy quark double differential collision rate with respect to final momentum $p'$ and scattering angle $\vartheta$, $\Gamma_2(p',\vartheta):=\partial^2\Gamma/(\partial p'\partial\vartheta)$. The rates are (pre-)computed in a thermalized QGP at rest; flow of the medium is taken into account by Lorentz boosts in the local rest frame\footnote{The rates are adjusted to the elapsed time in the local rest frame.}. The divergence from the $t$-channel terms in the Born cross sections is screened by the effective screening mass $\mu^2(t)$, and the running of the QCD coupling is taken into account as described in Section \ref{chap:200}.
Discretizing $p'$ and $\vartheta$, and multiplying $\Gamma_2(p',\vartheta)\,\Delta p' \Delta\vartheta$ by a sufficiently small time interval $\Delta t$ yields, for given initial momentum $p$, probability matrices describing transitions $p \to p'$ between momentum states with a deflection by $\vartheta$.
These stochastic matrices allow to interpret the sequence of heavy quark interactions as a Markov chain \cite{Peshier:2008bg}, and they form the basis of our Monte Carlo simulation which is an efficient alternative to other transport approaches such as \cite{Uphoff:2010sh}.

\begin{figure}
 \includegraphics[width=8.5cm,height=6.1cm]{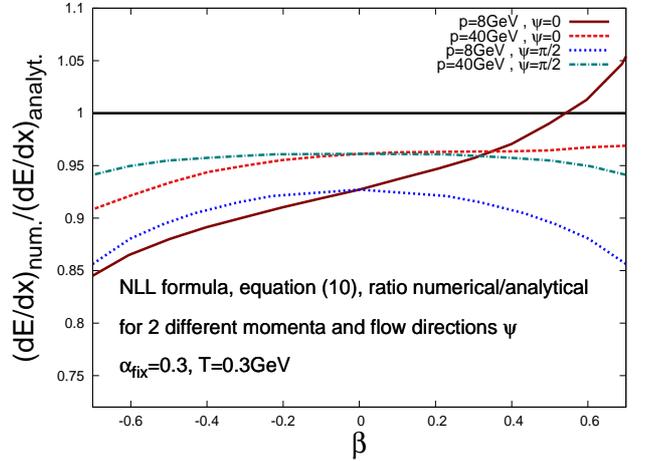}
 \centering
 \caption{Comparison of numerical results for $dE/dx$ to the analytical formula \eqref{eqn:216}. The numerical calculations were done within our MC-simulation for charm quarks with $p=8\,$GeV and $p=40\,$GeV.}
 \label{fig:301}
\end{figure}

The resulting (numerical) mean energy loss is compared in Fig.~\ref{fig:301} to the analytic formula \eqref{eqn:216}. This comparison illustrates that sub-leading terms in $E$, which are not captured by \eqref{eqn:216}, are reasonably small already for intermediate charm energies. It also shows that a potential dependence of the constant $c(n_f)$ on the flow velocity $\beta$ is suppressed at large energy, as anticipated in Section \ref{chap:200}.

In our approach the heavy quarks are treated as test particles which do not, in turn, affect the medium. This approximation is justified because of the small number of heavy flavor pairs, about 4 comparing to 800 gluons in the mid-rapidity region of a central collision at RHIC \cite{Uphoff:2010sh}.
This allows us to use different approaches, such as hydrodynamics or BAMPS (Boltzmann Approach of Multi-Parton Scattering \cite{Xu:2004mz}), to specify the temperature and flow profile of the QGP background in heavy-ion collisions. We note in this context that we have successfully tested our approach for various background dynamic scenarios by comparing to original BAMPS calculations (to that end we recalculated our interaction rates $\Gamma_2$ with the classical Boltzmann distribution, as assumed in BAMPS).

The spacial initialization of heavy quark dijet events is done according to the Glauber model \cite{Glauber:2006gd,Miller:2007ri} and the initial momentum distributions are computed with the event generator $MC@NLO$ at next-to-leading order precision \cite{Frixione:2002ik,Frixione:2003ei}. Shadowing and the Cronin effect are neglected since the modification of heavy quark distributions should be small for momenta $p_T\gsim 2\,$GeV \cite{Armesto:2005mz}. An indirect comparison of the initial momentum distributions to experimental data can be found in \cite{Uphoff:2011ad}.

When at a later stage in our simulation the QGP reaches the critical temperature $T_c = 200\,$GeV, we assume that the heavy quarks fragment into heavy mesons, described by the Peterson function \cite{Peterson:1982ak}
\be
D_Q^M\left(z\right)=\frac{N}{z\left[1-\frac{1}{z}-\frac{\epsilon_Q}{1-z}\right]}\,,
\label{eqn:301}
\ee
where $N$ is a normalization constant, $z=p_M/p_Q$ the momentum fraction (heavy meson to heavy quark) and $\epsilon_c=0.05$, $\epsilon_b=0.005$ are typical squared mass ratios for charm and bottom, respectively. The following stage of heavy meson decays to electrons is calculated with PYTHIA 8.1 \cite{Sjostrand:2006za,Sjostrand:2007gs}.

\section{Heavy-ion collisions}
\label{chap:400}
In order to study thoroughly the mechanisms which affect the spectrum of electrons from heavy quarks in heavy-ion collisions -- energy loss at parton level, hadronization to mesons and their subsequent decay -- we consider first the simple Bjorken model before investigating more realistic dynamics of the QGP background (in so doing we generalize the exploratory study of \cite{Peshier:2008bg}, which considered only the charm sector at partonic level). Our observable is the nuclear modification factor, defined as the ratio of the final transverse momentum distribution to the distribution in a nucleon-nucleon collision scaled by the number of binary collision $N_{b. coll}$ according to the Glauber assumption,
\be
R_{AA}=\frac{d^2N_{AA}/(dp_tdy)}{N_{b. coll}\, d^2N_{NN}/(dp_tdy)}\,.
\label{eqn:401}
\ee
We improve the simple power-law momentum spectrum assu\-med in \cite{Peshier:2008bg} by the NLO initial momentum distribution. Besides that, we use the same parameter sets for the dynamics/geometry (summarized in Table \ref{tab:401}), except for the hadronization temperature which is set to a slightly larger value, $T_c=200\,$MeV (which somewhat reduces the life time $\tau_c = \tau_0 T_0^3/T_c^3$ of the QGP phase), for consistency with the more detailed simulations below.
\begin{table}
\centering
\small{
\begin{tabular}{|c|c|c|c|c|c|}
\hline
 & $T_0\,|\,T_c\,[GeV]$ & $\tau_0\,|\,\tau_c\,$[fm/c] & $R\,[fm]$ & $dN_{ini}/dp_t^2$ \\
\hline
\hline
$set-1$ & $0.42\,|\,0.2$ & $0.6\,|\,5.6$ & $5.0$ & $NLO$ \\
\hline
$set-2$ & $0.30\,|\,0.2$ & $1.0\,|\,3.4$ & $6.6$ & $NLO$\\
\hline
\end{tabular}}
\caption{Parameter sets for the Bjorken model, $R$ denotes the effective radius of the nucleus.}
\label{tab:401}
\end{table}
\begin{figure}
 \includegraphics[width=8.5cm,height=6.1cm]{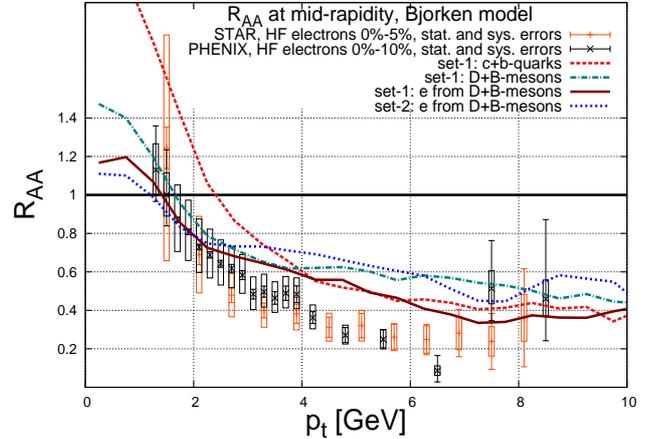}
 \centering
 \caption{Comparison of $c+b$, $D+B$ (Peterson fragmentation) and electron (decay from $D+B$ with PYTHIA) quenching at mid-rapidity in the Bjorken model to the observed $R_{AA}$ from heavy flavor decays in the centrality class of $0\%-10\%$ \cite{Abelev:2006db,Adare:2010de}.}
 \label{fig:401}
\end{figure}
Fig.~\ref{fig:401} shows the combined modification factors of $c+b$ quarks, $D+B$ mesons and the resulting electron yields. It is apparent (as assumed in \cite{Peshier:2008bg}) that the $c+b$ spectra are in good agreement with the electron spectra from heavy flavor decays for $p_t\gsim 4\,$GeV.\footnote{For smaller momenta the results should be considered with some caution since the cut-off parameter \eqref{eqn:207} was determined for large momenta, and because the fragmentation concept becomes questionable, shadowing needed to be taken into account etc.}
In this simplified framework we capture the trend of the experimental data, with some room for effects due to heavy quark radiative energy loss. However, since the radial expansion of the QGP will lead to shorter life times of the plasma phase, we now turn to more realistic dynamical models.

\begin{figure}[hbt]
 \includegraphics[width=8.5cm,height=6.1cm]{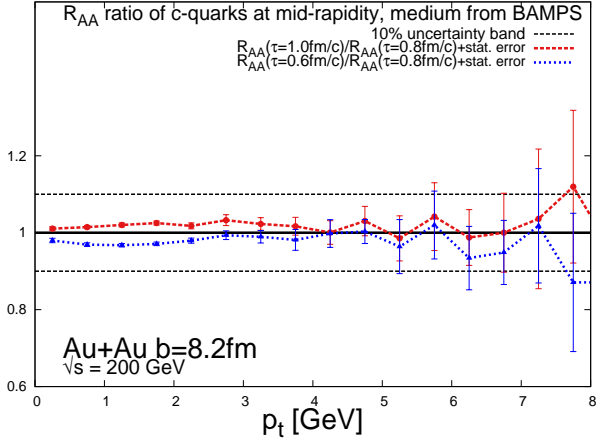}
 \centering
 \caption{(In)\,sensitivity of $R_{AA}$ in a peripheral nucleus-nucleus collision to different thermalization times $\tau = 0.6 \ldots 1.0\,$fm/c, which are used as starting point of our MC simulation.}
 \label{fig:402}
\end{figure}

The local temperature $T(\vec x,t)$ and the flow velocity $\vec \beta(\vec x,t)$ of the QGP background can be obtained from available parton transport codes or hydrodynamics. We focus here on the BAMPS approach \cite{Xu:2004mz}, which can reproduce short thermalization times of $\tau_0 = 0.6 \ldots 1\,$fm/c as expected from experimental data \cite{Xu:2004mz,Heinz:2004pj} (and which were assumed for the parameter sets in Table~\ref{tab:401}).
We demonstrate with Fig.~\ref{fig:402} that the quenching ratio is not very sensitive to the value of $\tau_0$.

In order to control that the observables under consideration are not very sensitive to other details of the background dynamics, we also use temperature and flow profiles as computed with the (3+1)-dimensional ideal hydrodynamic solver SHASTA \cite{Rischke:1995ir}. Here we set the initial energy density to that of BAMPS and assume an initialization time of $0.6 \ldots 0.8\,$fm/c.

\begin{figure}
 \includegraphics[width=8.5cm,height=6.1cm]{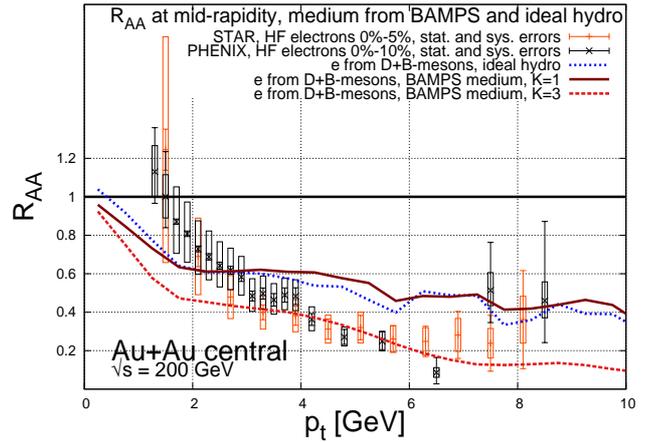}
 \centering
 \caption{Nuclear modification factor $R_{AA}$ at mid-rapidity $|y|<0.35$ for a central $Au+Au$ collision with a center of mass energy of $\sqrt{s}=200\,$GeV (RHIC). The curves are calculated within our transition matrix approach for the NLO initial momentum distribution of heavy quarks. $K$ denotes an artificial scaling factor of the elastic cross section. For comparison, experimental data on electrons from heavy flavor decays in the centrality class of $0\%-10\%$ are shown \cite{Abelev:2006db,Adare:2010de}.}
 \label{fig:403}
\end{figure}

\begin{figure}
 \includegraphics[width=8.5cm,height=6.1cm]{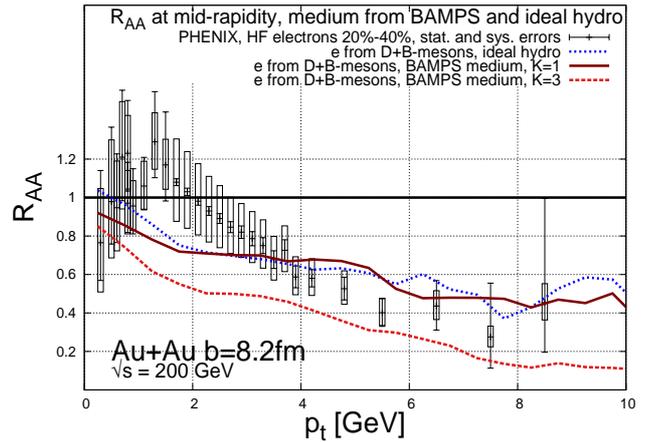}
 \centering
 \caption{Same as Fig.~\ref{fig:403} but for a peripheral $Au+Au$ collision with an impact parameter of $b=8.2\,$fm. For comparison, experimental data on electrons from heavy flavor decays in the centrality class of $20\%-40\%$ are shown \cite{Adare:2010de}.}
 \label{fig:404}
\end{figure}

Figs.~\ref{fig:403} and \ref{fig:404} show our results for the nuclear modification factor in a central and in a peripheral collision with an impact parameter of $b=8.2\,$fm, which corresponds to the centrality class of $20\%-40\%$. In both cases the quenching due to collisional energy loss alone reproduces the trend of the data, again with some room for radiative effects, although somewhat less for more peripheral collisions. Following a common attempt to estimate the relative importance of collisional vs.\ radiative energy loss we scale {\em ad hoc} the elastic cross section by a factor $K$. We note that this prescription can at best give indications since both the energy and the path length dependence of the two loss mechanisms differ markedly. It turns out that the experimental data for both central and peripheral collisions lead to a constraint $1 \ksim K \ksim 3$, from which we may conclude that collisional and radiative processes are of similar importance. We note that our $K\approx 2-3$ (from peripheral to central collisions) is somewhat smaller than $K_{\rm BAMPS} \approx 4$ found in \cite{Uphoff:2011ad}, which can be traced back to BAMPS simulating only gluons distributed according to Boltzmann statistics (with the effect of the light-flavor degrees of freedom estimated by color-factor scaling).

After the nuclear modification factor $R_{AA}$ we now consider the elliptic flow, which is defined by the ensemble average
\be
v_2(p_t) = \left<\frac{p_x^2-p_y^2}{p_t^2}\right>\,.
\label{eqn:402}
\ee
The elliptic flow of heavy quarks forms at later stages of heavy-ion collisions by interactions with the medium; consequently the sensitivity of the following results on the thermalization time $\tau_0$ is even less than for $R_{AA}$.

\begin{figure}
 \includegraphics[width=8.5cm,height=6.1cm]{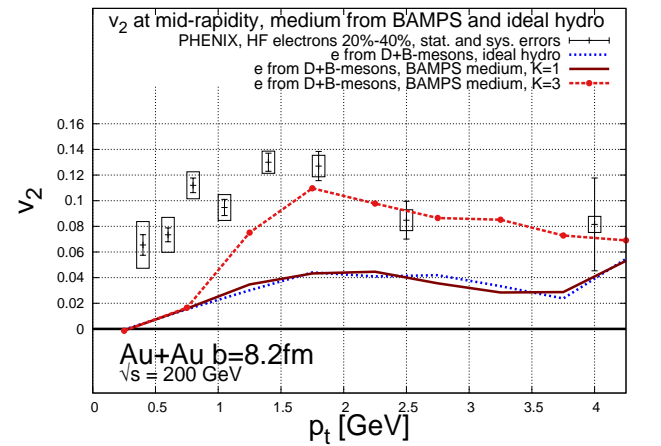}
 \centering
 \caption{Same as Fig.~\ref{fig:404} but for elliptic flow.}
 \label{fig:405}
\end{figure}

Fig.~\ref{fig:405} shows the transverse momentum dependence of $v_2$ obtained in our approach in comparison with experimental data.
Obviously, the data are underestimated considerably; for a better agreement the binary cross sections would need to be multiplied by a factor of $K^{v_2} \approx 3$, which is slightly larger than what we found for $R_{AA}$, but again less than $K_{\rm BAMPS}^{v_2} \approx 4$ inferred in the BAMPS approach \cite{Uphoff:2011ad}.
We emphasize here again that the crucial small value \eqref{eqn:207} of the cut-off parameter $\kappa$ in \eqref{eqn:206} is justified only in the high-energy limit $E \gsim M^2/T$, i.\,e.\ a few GeV for charm quarks \cite{Peigne:2007sd,Peigne:2008nd}. Lacking a similar calculation for a cut-off at smaller energies, we have extrapolated our approach to rather small $p_t$ values, therefore the results can only be considered as rough estimates.

\section{Conclusion}
\label{chap:500}
We have generalized the leading logarithmic result for the mean energy loss of energetic heavy quarks in the QGP to the situation of a constant flow of the medium. The result has been tested numerically using our transition matrix approach, which incorporates running coupling and uses an improved infrared regulator, and which allows us compute efficiently the evolution of heavy quarks in any thermalized medium of quarks and gluons due to binary collisions.

As an application we have calculated the heavy flavor nuclear modification factor and the elliptic flow in heavy-ion collisions. While binary scattering of heavy quarks in the QGP alone cannot explain the experimental data, this mechanism contributes significantly, of a similar order as radiative processes.
This can be quantified phenomenologically by a parameter $K$ multiplying the elastic cross sections. In case of the nuclear modification factor we would need an increase $K \approx 2-3$, whereas matching better the elliptic flow data requires $K^{v_2} \approx 3$; both values are somewhat smaller than findings in other approaches, basically due to an improved infrared cut-off and using running coupling and full quantum statistics.

The fact that a simple upscaling of the binary cross sections seems not sufficient to reproduce $R_{AA}$ and $v_2$ data simultaneously may help to understand better the energy loss due to strongly anisotropic gluon emission (see \cite{Abir:2011jb,Abir:2012pu}) and/or the initial conditions of the medium, which could modify its transverse expansion. Furthermore our results for the $R_{AA}$ of central and peripheral collisions may be a direct indication of non-linear scaling in the pathlength of the radiative energy loss. Since the mean pathlength drops in non-central collisions this point of view explains naturally a small increase in $K$ from peripheral to central collisions in the case of $R_{AA}$. A higher value of $K^{v_2}$ for the elliptic flow in comparison to $K$ of the nuclear modification factor could also be explained due to a different pathlength dependence of the radiative scenario, which amplifies the path asymmetry in and out of plane in a peripheral collision.
On the other hand, it seems also necessary to improve our understanding of elastic scattering processes in the low-$p_t$ region.
\\\\\\\\
\textbf{Acknowledgments:}

A.\ M.\ and J.\ U.\ acknowledge the kind hospitality at the University of Cape Town, where part of this work has been done. Part of the calculations was performed at the Center for Scientific Computing of the Goethe University Frankfurt. This work was supported by the Helmholtz International Center for FAIR within the framework of the LOEWE program launched by the State of Hesse, and by the South African National Research Foundation.

\appendix
\section{Heavy-quark energy loss in a streaming medium}
\label{appendix:100}
We derive here the Coulomb contribution in formula \eqref{eqn:216} for the mean energy loss in the presence of a collective QGP flow (neglecting the momentum dependence of the QCD coupling, see footnote \ref{footnote: run coupling}). As for the medium at rest, the leading logarithm results from terms $d\sigma^{[t]}/dt \sim 1/t^{2}$ in the cross section, and we can extract it from the hard contribution \eqref{eqn:204} with $t^\star$ replaced by $-m_D^2$.
For large energy, we can also simplify $t_{\rm min} \simeq -s$, hence (omitting prefactors, which can be fixed by matching to the static case)
\be
	\frac{dE^{[t]}}{dx}
	\sim
	\int_k\frac{n_J\left(\vec k\right)}{2k}\ln\left(\frac{s}{m_D^2}\right)\, ,
\label{eqn:a1}
\ee
where $n_J\left(\vec k\right)$ denotes the J\"uttner distribution function \eqref{eqn:208}.

To perform the $k$-integral, it is convenient to specify a coordinate system. In spherical coordinates, with the heavy quark momentum $\vec p$ in $z$-direction, the flow velocity $\vec \beta$ in the $yz$-plane, viz.\ $\vec p\cdot\vec\beta = p\beta\cos\psi$, and the orientation of the thermal momentum $k$ specified by the polar angle $\theta$ and azimuth $\phi$, the argument of the J\"uttner function reads
\be
	\frac{k_\mu\beta^\mu}{T}
	=k \frac{\gamma}{T}\left[1-\beta\left(\cos\theta\cos\psi+\sin\theta\sin\psi\cos\phi\right)\right]=\frac{k}{T_{\rm eff}}\,,
\label{eqn:a2}
\ee
where we introduce an effective temperature.
Approximating $s \simeq 2Ek(1-x)$ with $x := \cos\theta$, we rewrite equation \eqref{eqn:a1},
\be
	\frac{dE^{[t]}}{dx}
	\sim
	\int_{-1}^1 dx\int_0^{2\pi} d\phi\int_0^\infty dk k\, n_J\left(\vec k \right)\ln\left(\frac{2Ek(1-x)}{m_D^2}\right),
\label{eqn:a3}
\ee
and integrate over $k$,
\bean
&&\int_0^\infty dk k\, n_J\left(\vec k\right)\left[\ln\left(\frac{ET_{\rm eff}(1-x)}{m_D^2}\right)+\ln\left(\frac{2k}{T_{\rm eff}}\right)\right]
\\
&&\!\!\simeq\; T_{\rm eff}^2\ln\left(\frac{ET_{\rm eff}}{m_D^2}(1-x)\right)\,,
\eean
where we have neglected terms which will not contribute to the logarithm in $E$. We proceed with the $\phi$-integral in \eqref{eqn:a3}, by writing the effective temperature introduced in \eqref{eqn:a2} as $T_{\rm eff} = T/(\gamma r)$, with $r = A-B\cos\phi$ where $A = 1-\beta x\cos\psi$ and $B = \beta\sqrt{1-x^2}\sin\psi$,
\bean
&&\frac{T^2}{\gamma^2}\int_0^{2\pi}d\phi\,
\frac1{r^2}\left[\ln\left(\textstyle\frac{ET}{\gamma m_D^2}(1-x)\right)-\ln(r)\right]
\\
&&\!\!=\; \frac{T^2}{\gamma^2}\left[\ln\left(\textstyle\frac{ET}{\gamma m_D^2}(1-x)\right)I_1-I_2\right]\,.
\eean
The auxiliary integrals $I_1$ and $I_2$ can be calculated in closed form,
\[
\begin{aligned}
I_1=&\int_0^{2\pi}\frac{d\phi}{2\pi}\frac{1}{r^2}=\frac{A^{-2}}{(1-C^2)^{3/2}}\,,\\
I_2=&\int_0^{2\pi}\frac{d\phi}{2\pi}\frac{\ln\left(r\right)}{r^2}\\
=&\;\frac{A^{-2}}{(1-C^2)^{3/2}}\Bigg[\ln\left(A\right)\\
&+\ln\left(2\frac{1-C^2}{C^2}\left(1-\sqrt{1-C^2}\right)+\sqrt{1-C^2}-1\right)\Bigg]\,,
\end{aligned}
\]
where $C = B/A = \beta\sqrt{1-x^2}\sin\psi/\left(1-\beta x\cos\psi\right)$. The final $x$-integration in \eqref{eqn:a3} can then be performed using
\[
\begin{aligned}
\int_{-1}^1 dx\,I_1&\;=\;2\gamma^2\,,\\
\int_{-1}^1 dx\,I_1\ln\left(1-x\right)&\;=\;2\gamma^2\left[\ln\left(1-\beta\cos\psi\right)+\ln\left(2\gamma\right)
+\frac{L}{2\beta}\right]\,,\\
\int_{-1}^1 dx\,I_2&\;=\;2\gamma^2\left[1-\ln\left(\gamma\right)+\frac{L}{2\beta}\right] ,
\end{aligned}
\]
where $L := \ln\left(\frac{1-\beta}{1+\beta}\right)$.
Collecting all relevant terms contributing to the energy loss due to $t$-channel scattering at leading logarithmic order with flow, we arrive at
\[
	\frac{dE^{[t]}}{dx}
	\sim T^2\ln\left(\frac{\gamma E\left(1-\beta\cos\psi\right) T}{m_D^2}\right),
\]
where we indeed recognize in the argument of the logarithm the 4-product $p_\mu\beta^\mu$, as anticipated in Eq.~\eqref{eqn:216}. The calculation of the collinear logarithmic contribution can be done along similar lines.



\end{document}